\documentstyle[aps,prb,tighten,multicol,epsfig]{revtex} 

\begin{document}

%\draft   

\title{Numerical study of quasiparticle lifetime in quantum dots}

\author{Alejandro M. F. Rivas and Eduardo R. Mucciolo}

\address{Departamento de F\a'{\i}sica, Pontif\a'{\i}cia Universidade
Cat\'olica do Rio de Janeiro, CP 38071, 22452-970 Rio de
Janeiro, Brazil}

\author{Alex Kamenev}

\address{Department of Physics, Technion, Haifa, 32000, Israel}

\date{September 25, 2001}

\maketitle

%twocolumn

%%%%%%%%%%%%%%%%%%%%%%%%%%%%%%%%%%%%%%%%%%%%%%%%%%%%%%%%%%%%%%%%%%%%%%%%%%%%%%
\begin{abstract}

The decay rate of quasiparticles in quantum dots is studied through
the real time calculation of the single-particle Green function in the
self-consistent approximation. The method avoids exact
diagonalization, transforming the problem into a system of coupled
non-linear integral equations which may be solved iteratively. That
allows us to study systems larger than previously treated in the
literature. Our results for the inverse participation ratio of
many-body states show that the threshold energy for the quasiparticle
disintegration is $E^\ast \sim \sqrt{g} \Delta$. The delocalization
transition is soft rather than sharp. Three different regimes as
function of the effective interaction strength may be clearly
identified at high energies.
\end{abstract}
%%%%%%%%%%%%%%%%%%%%%%%%%%%%%%%%%%%%%%%%%%%%%%%%%%%%%%%%%%%%%%%%%%%%%%%%%%


% 71.10.-w Theories and models of many electron systems
% 73.23.-b Mesoscopic systems

%%%%%%%%%%%%%%%%%%%%%%%%%%%%%%%%%%%%%%%%%%%%%%%%%%%%%%%%%%%%%%%%%%%%%%%%%%

\begin{multicols}{2}

%\narrowtext

%%%%%%%%%%%%%%%%%%%%%%%%%%%%%%%%%%%%%%%%%%%%%%%%%%%%%%%%%%%%%%%%%%%%%%%%%%

\section{Introduction}

%%%%%%%%%%%%%%%%%%%%%%%%%%%%%%%%%%%%%%%%%%%%%%%%%%%%%%%%%%%%%%%%%%%%%%%%%%

Isolated quantum dots provide a unique opportunity to study the
interplay between phase coherence, disorder, and electron-electron
interaction effects in a confined geometry. These systems are realized
experimentally in several forms; the best known example are the
electron islands formed by the electrostatic lateral confinement of a
two-dimensional electron gas in GaAs/GaAlAs
heterostructures.\cite{dotsrevision} Among the important issues
recently investigated is the lifetime of quasiparticle excitations
that can be associated with the broadening of tunneling conductance
peaks. In experiment reported in Ref. \onlinecite{sivan94}, for
instance, few narrow peaks were observed above the Fermi energy,
followed by a broad continuum. This result has initiated a theoretical
discussion about the nature of the excited states in confined (but
large) fermionic systems.

In this paper we study numerically the problem of quasiparticle decay
rate in finite fermionic systems with chaotic single-particle
dynamics. In most quantum dots, the presence of either disorder or
complex boundaries lead to single-particle eigenstates displaying
universal statistical properties. The universality takes place for
states within the energy range of order $g\Delta$ near the Fermi
energy, where $g \gg 1$ is the dimensionless conductance of the system
(i.e, conductance in units $e^2/\hbar$) and $\Delta$ is the mean
spacing between neighboring single-particle eigenenergies. The
characteristic features of such states are level repulsion and
spectral rigidity.\cite{mehta91} Moreover, the two-body (short-ranged)
interaction matrix elements taken in the single-particle eigenbasis
fluctuate according to Gaussian distribution whose standard deviation
scales as $1/g$.\cite{mirlin99}

Ref. \onlinecite{altshuler97} proposed to map the problem of
quasiparticle lifetime onto an Anderson problem in the many-body Fock
space. Upon such mapping, many-body states of the noninteracting
system are considered as ``sites'' of the Anderson lattice, while the
interactions provide hopping between them. Since both the ``on-site''
energies and the hopping matrix elements are random (disorder
dependent), one may expect a localization-delocalization transition to
take place for certain values of the system parameters. For
excitations with energy $\varepsilon$ lower than the threshold energy
$E^\ast$, the exact many-body states are ``localized'' in the
single-particle basis. As a result, an excited quasiparticle evolves
into a superposition of only few low-energy many-body eigenstates. The
spectral function consists of a finite number of $\delta$-functional
peaks and the quasiparticle lifetime is essentially infinite in this
regime. At the energies above the delocalization threshold,
$\varepsilon > E^\ast$, a large number of basis single-particle states
contribute to a given exact many-body states. This eventually leads to
spectral functions of the Breit-Wigner form whose width (inverse
lifetime) may be evaluated employing the golden rule (GR).

To work out details of this crossover, Ref. \onlinecite{altshuler97}
employed an approximate mapping of the many-body space of a fermionic
system onto a Cayley tree (CT). Each generation of the CT corresponds
to a set of many-body states with a given number of electron-hole
pairs. Interactions (hopping) provide coupling between generations by
exciting new electron-hole pairs. The advantage of such analogy is
that the Anderson localization problem on the CT may be solved
exactly.\cite{abou-chacra73,mirlin97} Such solution indicates the
existence of a sharp crossover at the energy $E^\ast \sim \Delta
\sqrt{g/\ln g}$, where the participation ratio $P$ jumps abruptly from
a value close to unity in the localized phase ($\varepsilon < E^\ast$)
to infinity in the delocalized phase ($\varepsilon > E^\ast$). The
other important feature of the exact CT solution is the existence of a
second characteristic energy $E^{\ast\ast} \sim \Delta \sqrt{g} >
E^\ast$. For excitation energies $\varepsilon > E^{\ast\ast}$, the
spectral density is of the Breit-Wigner form and the GR applies. On
the other hand, for energies $E^\ast < \varepsilon < E^{\ast\ast}$,
the spectral density consists of a very large number of narrow, but
separate peaks (though the participation ratio is infinite).

The disadvantage of the CT mapping is that it is by no means
exact. Indeed, it disregards potentially important effects such as
existence of loops in the many-body space, correlations between
on-site energies of different generations, and the fact that the
actual number of accessible many-body states at a given energy is
finite. Therefore, the predictions of the exact CT solution should be
taken with a grain of salt. In fact, it is clear that no abrupt
transition may happen, since the number of many-body states at the
energy close to the critical one is exponentially large, but
finite. As a result, the very existence of the transition, its
quantitative features, and the characteristic energy scales are not
firmly established by the CT picture. Several theoretical and
numerical
works\cite{jacquod97,mirlin97,silvestrov97,berkovits98,georgeot97,mejia-monasterio98,leyronas99,silvestrov98,leyronas00}
were devoted to clarification of these issues. One numerical
study\cite{mejia-monasterio98} claimed that for small systems the
transition is absent and the quasiparticle lifetime decay can be well
described by the GR. That conclusion does not apply to larger systems,
where a breakdown of the GR is observed \cite{leyronas99}. In large
systems, one can explore a broader range of values of $g$, as well as
higher excitation energies. Nevertheless, no signal of an abrupt
transition was found. Essentially, all numerical work so far has
relied on the exact diagonalization of the full many-body Hamiltonian
(thus limited size Fock space), although under different assumptions
regarding the structure of the two-body matrix elements.

Here we employ an alternative approach suggested by
Levitov.\cite{Levitov} The method avoids matrix inversion or
diagonalization and allows the investigation of large size systems. It
is based on the calculation of the single-particle Green's function in
real time. Adopting the self-consistent approximation for the
self-energy, one can transform the problem into a non-linear integral
equation which may be solved iteratively. The inverse participation
ratio that measures the overlap between a one-particle state and the
many-body eigenstates may be obtained directly from the long-time
behavior of the Green's function. Our simulations involved up to 50
particles and 150 basis states (thus with filling factor $1/3$; we
deal with the spinless fermions for simplicity). The results confirm
conclusions of Ref. \onlinecite{leyronas99}, namely, that there is no
sharp transition of the participation ratio (contrary to the case of
an ideal CT). However, our estimate for the threshold energy at which
quasiparticle states delocalize does not match the prediction of
Ref. \onlinecite{altshuler97}. Instead, our results are consistent
with the scaling of the threshold energy as $\Delta\sqrt{g}$. We see
no indications of other scalings such as those predicted in
Ref. \onlinecite{silvestrov98}.

At the same time, our numerical data is consistent with the existence
of the three different energy regimes, in agreement with
Ref. \onlinecite{altshuler97}. At small energies, single-particle
states overlap with only a finite number of many-body states and
quasiparticles do not decay. In the the opposite limit, of large
energy, the quasiparticles decay rapidly into the continuum of
many-body states. The decay rate follows closely the GR prediction. In
the intermediate energy range, the quasiparticle is neither well
defined (Fock-space localized), nor short lived, and GR does not
apply.

This paper is organized as follows. In Sec. \ref{sec:II} we formulate
the method used in the numerical simulations. The results are
presented in Sec. \ref{sec:III}. Discussion and conclusions are left
to Sec. \ref{sec:IV}.

%%%%%%%%%%%%%%%%%%%%%%%%%%%%%%%%%%%%%%%%%%%%%%%%%%%%%%%%%%%%%%%%%%%%%%%%%%

\section{Real-time Green's function approach}
\label{sec:II}

%%%%%%%%%%%%%%%%%%%%%%%%%%%%%%%%%%%%%%%%%%%%%%%%%%%%%%%%%%%%%%%%%%%%%%%%%%

The Hamiltonian of an isolated quantum dot with $N$ spinless electrons
may be separated into two parts,
\begin{equation}
H = H_0 + H_1,
\end{equation}
where $H_0$ contains one-body terms, such as the kinetic energy and
the background potential, and $H_1$ represents the two-body
interactions. We denote $\varepsilon_j$ and $\phi_j(x)$ the
single-particle eigenvalues and eigenfunctions of $H_0$,
respectively. In this eigenbasis, one may write
\begin{equation}
H_0 = \sum_j \varepsilon_j\, c_j^\dagger c_j
\end{equation}
and
\begin{equation}
H_1 = \sum_{j<k,m<l} V_{jk}^{lm}\, c^\dagger_m c^\dagger_l c_j c_k,
\end{equation}
where the fermionic operators $c_j$ and $c_j^\dagger$ are,
respectively, annihilation and creation operators of an electron on
the state $j$. The interaction matrix elements are given by
\begin{equation}
V_{jk}^{lm} = \int\! dx \int\! dx^\prime\, \phi_m^\ast(x)
\phi_l^\ast(x^\prime)\, V(x-x^\prime)\, \phi_k(x) \phi_j(x^\prime).
\end{equation}
Below, we consider the case of short-range interactions of the form
\begin{equation}
V(x) = \lambda\, \Delta\, L^d\, \delta^{(d)}(x)\, ,
\end{equation}
where $L^d$ is the system volume and $\lambda$ is the dimensionless
interaction strength. Diagonal and off-diagonal matrix elements of the
interaction have distinct statistical properties.\cite{altshuler97}
The diagonal ones,
\begin{equation}
V_{jk}^{jk} = V_{jk}^{kj} = \lambda \Delta\, L^d \int\! dx\, |\phi_k
(x)|^2 |\phi_j (x)|^2 ,
\end{equation}
have non-zero average value, $ \overline{V_{jk}^{jk}} \sim
\lambda\Delta$, while fluctuations around this mean value are small as
$g^{-1}$. On the other hand the off--diagonal matrix elements
\begin{equation}
V_{jk}^{lm} = \lambda \Delta\, L^d \int dx\, \phi_m^\ast (x)
\phi_l^\ast (x) \phi_k (x) \phi_j (x) 
\end{equation}
are zero in average, $\overline{ V_{jk}^{lm} } = 0$, while their
typical value is $\overline{ |V_{jk}^{lm}| } = \lambda\Delta/g$ for
any choice of different $j,k,l,m$ within the universal range $\Delta
g$.\cite{mirlin99}

In the absence of $H_1$, the ground state of the system is formed by
the Slater determinant of the $N$ lowest energy single particle
eigenstates. That is, for eigenvalues arranged in ascending order, all
states below and including the Fermi energy $\varepsilon_F =
\varepsilon_N$ are occupied and the remaining are empty. Excitations
are formed by promoting one or more particles from the Fermi sea to
the empty states above $\varepsilon_F$, leaving the holes behind. For
instance, for an excited eigenstate formed by $m$ particle-hole pairs,
\begin{eqnarray}
| j_{2m},\ldots,j_{m+1};j_m,\ldots,j_1 \rangle_0 & = &
 c^\dagger_{j_{2m}} c^\dagger_{j_{m+1}} \cdots \nonumber \\ & & \times
 \, c_{j_m} \cdots c_{j_1} | {\rm FS} \rangle_0.
\end{eqnarray}
(Here, the subscript $0$ is used to stress the absence of
interactions.)

%%%%%%%%%%%%%%%%%%%%%%%%%%%%%%%%%%%%%%%%%%%%%%%%%%%%%%%%%%%%%%%%%%%%%%%%%%
\subsection{The single--particle Green's function}

Let us consider the (zero-temperature) time-ordered, single-particle
Green's function ($\hbar =1$):
\begin{equation}
G_j(t) \equiv - i \left\langle {\rm FS} \left| T \left\{ c_j(t)\,
c_j^\dagger(0) \right\} \right|{\rm FS} \right\rangle ,
\label{eq:GF}
\end{equation}
where $|{\rm FS} \rangle$ is the exact many-body ground state of the
full Hamiltonian and $T$ denotes the time ordering.

It is useful to review briefly the non-interacting case. The Green's
function can then be written in the following form,
\begin{equation}
G_j^{(0)}(t) = \left\{ \begin{array}{lr} -i \theta(t)\,
e^{-i\varepsilon_j t}, & \varepsilon_j > \varepsilon_F \\ i
\theta(-t)\, e^{-i\varepsilon_j t}, & \varepsilon_j \le \varepsilon_F
\end{array} \right. ,
\label{eq:zerothGF}
\end{equation}
where $\theta(t)$ is the Heaviside step function. It is
straightforward to show that
\begin{equation}
(i\partial_t - \varepsilon_j) G_j^{(0)}(t) = \delta(t),
\label{eq:Eq0}
\end{equation}
with the initial conditions
\begin{eqnarray}
G_j^{(0)}(0^+) & = & -i, \qquad \varepsilon_j > \varepsilon_F^{(0)} \\
G_j^{(0)}(0^-) & = & i, \qquad \varepsilon_j \le \varepsilon_F^{(0)}
\end{eqnarray}
After the Fourier transform of Eq. (\ref{eq:Eq0}), one finds
\begin{equation}
\tilde{G}^{(0)}_j(E) = \frac{1}{E -\varepsilon_j + i0^+\,
\mbox{sgn}(\varepsilon_j)},
\end{equation}
Thus, $\tilde{G}^{(0)}_j(E)$ has a pole whose real part is exactly the
$j$-th single-particle eigenenergy.

In the presence of interactions, $H_1$, the ground state is a
superposition of a large number of Slater determinants. The same
applies for the excited states, which are formed by a superposition of
many particle-hole states. Nevertheless, many aspects of the
interactions, not related to the correlation effects, may be taken
into account in the Hartree-Fock approximation. For instance, the
diagonal part of $H_1$ (related to matrix elements $V_{jk}^{jk}$ and
$V_{jk}^{kj}$) can be incorporated into $H_0$, creating new sets of
single-particle eigenstates $\{ \phi_j^{\rm HF} \}$ and eigenenergies
$\{ \varepsilon_j^{\rm HF} \}$, with the statistical properties
similar to the noninteracting ones.\cite{Orgad99} For our purposes, it
is unnecessary to distinguish this new set from the original one and
the Hartree-Fock superscript is dropped. By the same token, one may
drop the diagonal elements. The off-diagonal ones are kept and are
solely responsible for the finite lifetime of the quasiparticles.

The exact single-particle Green's function may be expanded
perturbatively in ascending powers of $\lambda$. Through standard
diagrammatic techniques, one arrives at the Dyson equation which
generalizes Eq. (\ref{eq:Eq0}) onto the interacting case,
\begin{equation}
(i\partial_t - \varepsilon_j)\, G_j(t) = \delta(t) + \int\! dt^\prime\
\Sigma_j(t-t^\prime)\, G_j(t^\prime).
\label{eq:dyson}
\end{equation}
The self-energy $\Sigma_j(t)$ is the sum of all irreducible diagrams
and contains the information about the structure of the excited
states. In the frequency representation, one obtains
\begin{equation}
\tilde{G}_j(E) = \frac{1}{E -\varepsilon_j - \tilde{\Sigma}_j(E)}.
\end{equation}
Thus, the self-energy shifts the pole of the Green's function with
respect to the single-particle eigenenergy.

In the scheme proposed in Ref. \onlinecite{altshuler97}, the
self-energy involves only a limited class of diagrams. In order to
motivate the approximation involved, let us view each excited
many-body state formed by a Slater determinant of $N$ eigenstates
$\{\phi_j\}$ as a site in Fock space. The two-body interaction term
acts as a hopping matrix elements between sites, while the energy at
each site is formed by a sum of the energies of the single-particle
eigenstates present in the corresponding Slater determinant. The sites
are considered as nearest neighbors when they differ only by one
single-particle eigenstate. In fact, a more natural classification of
the excited many-body states uses the number of particle and hole
states.\cite{mejia-monasterio98} Taking the noninteracting Fermi sea
as a reference, we call a many-body state of class $m$ ($m>0$) when it
is formed by $m$ particle and $m-1$ hole excitations. In
Ref. \onlinecite{altshuler97}, hopping between sites of the same class
were not allowed; a site in class $m$ could connect to all accessible
sites in class $m+1$, but to only one site in class $m-1$. It is not
difficult to see that these restrictions lead to the topology of a CT.
It is important to notice that the coordination number of the tree at
a given site depends on the site class and the excitation energy;
eventually, for a sufficiently high class number, one runs out of
single-particle states and further branching is interrupted. This fact
was not properly taken into account in Ref. \onlinecite{altshuler97}
when applying the results of Anderson localization
problem.\cite{abou-chacra73} As a result, while the Fock space and the
CT localization problems share a similar mathematical structure, they
are not fully equivalent.

\begin{figure}
\epsfxsize=7cm\epsfbox{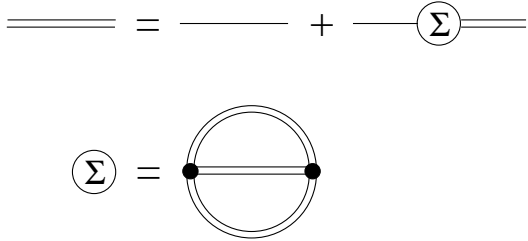} 
\caption{Dyson equation and self-energy.}
\label{fig:selfenergy}
\end{figure}

The same physics as in Ref. \onlinecite{altshuler97}, with the proper
branching number and on-site energy correlations, is fully encoded in
the following self-consistent approximation for the self-energy,
\begin{equation}
\Sigma_j(t) = \sum_{klm} \left| V^{jm}_{kl} \right|^2\, G_k(t)\,
G_l(t)\, G_m(-t).
\label{eq:selfenergy}
\end{equation}
The resulting self-consistent equation for the Green's function is
represented diagrammatically in Fig. (\ref{fig:selfenergy}). Equation
(\ref{eq:selfenergy}) for the self-energy is the main approximation
adopted in the present paper. We believe that it encompasses the
essential physics of the model. In particular, it includes the
``correct'' CT structure with floating coordination number, as well as
some of the possible loops in the many-body space. On the other hand,
we have to stress that it is certainly not exact and some of the
correlation effects are neglected by adopting
Eq. (\ref{eq:selfenergy}).

%%%%%%%%%%%%%%%%%%%%%%%%%%%%%%%%%%%%%%%%%%%%%%%%%%%%%%%%%%%%%%%%%%%%%%%%%%
 
%%%%%%%%%%%%%%%%%%%%%%%%%%%%%%%%%%%%%%%%%%%%%%%%%%%%%%%%%%%%%%%%%%%%%%%%%%

\subsection{The retardation ansatz}

For our purposes, it is useful \cite{Levitov} to work with the
real-time formulation, as in Eqs. (\ref{eq:dyson}) and
(\ref{eq:selfenergy}), instead of the energy representation. In order
to allow for an iterative solution of the non-linear coupled
equations, we decompose the time-ordered Green's in particle-like and
hole-like parts, depending on whether the index reference state $j$ is
above ($\varepsilon_j > \varepsilon_F$) or below ($\varepsilon_j \le
\varepsilon_F$) the Fermi energy, respectively. That is,
\begin{equation}
G_j(t) = \left\{ \begin{array}{lr} -i\, e^{-\varepsilon_j t}\, g_j(t),
& \varepsilon_j > \varepsilon_F \\ i \, e^{-\varepsilon_j t}\,
f_j(-t), & \varepsilon_j \le \varepsilon_F \end{array} \right. .
\label{eq:ansatz}
\end{equation}
It is now straightforward to obtain a set of integro-differential
equations for the functions $g_j(t)$ and $f_j(t)$. Let us, for
convenience, assume $\varepsilon_F = 0$, such that $j>0$ corresponds
to particle-like excitations, whereas $j \le 0$ corresponds to
hole-like excitations.

Using Eq. (\ref{eq:ansatz}) into Eqs. (\ref{eq:dyson}) and
(\ref{eq:selfenergy}) for the case $j > 0$, we find that
\begin{equation}
\frac{dg_j(t)}{dt} = \delta(t) - \int dt^\prime\, S_j(t-t^\prime)\,
g_j(t^\prime)
\label{eq:particleq}
\end{equation}
with the kernel
\begin{equation}
S_j(t) = \sum_{\stackrel{\scriptstyle k,l>0}{\scriptstyle m\le 0}}
\left| V^{jm}_{kl} \right|^2 e^{it(\varepsilon_j - \varepsilon_k -
\varepsilon_l + \varepsilon_m)} g_k(t) g_l(t) f_m(t).
\label{eq:particlekernel}
\end{equation}
Similarly, for $j \le 0$ we obtain
\begin{equation}
\frac{df_j(t)}{dt} = \delta(t) - \int dt^\prime\, R_j(t-t^\prime)\,
f_j(t^\prime)
\label{eq:holeq}
\end{equation}
with the kernel
\begin{equation}
R_j(t) = \sum_{\stackrel{\scriptstyle k,l\le 0}{\scriptstyle m> 0}}
\left| V^{jm}_{kl} \right|^2 e^{-it(\varepsilon_j - \varepsilon_k -
\varepsilon_l + \varepsilon_m)} f_k(t) f_l(t) g_m(t).
\label{eq:holekernel}
\end{equation}

So far no additional approximation to Eq. (\ref{eq:selfenergy}) have
been made. However, the self-consistent equations are still too
complicated to be treated numerically. To proceed, we can explore the
fact that the interactions are moderately weak ($\lambda \sim 1
$). Thus, the matrix elements $V^{jm}_{kl}$ mix strongly only states
that are nearly on shell, such that $\delta \varepsilon =
\varepsilon_j - \varepsilon_k - \varepsilon_l + \varepsilon_m$ is of
order of $\lambda \Delta /g$. Since we are interested in, say,
particle-like states with the typical energy $\varepsilon \gg
\delta\varepsilon$ we can disregard their admixture with the hole-like
states and vice-versa. Neglecting such mixing amounts to looking for
the solutions of the coupled integro-differential equations in the
retarded form. (As matter of fact we disregard the difference between
the retarded and time-ordered Green's functions for sufficiently large
energies.) We thus set the initial conditions as stated above and
substitute Eqs. (\ref{eq:particleq}) and (\ref{eq:holeq}) by
\begin{equation}
\frac{dg_j(t)}{dt} = - \int_0^t dt^\prime\, S_j(t-t^\prime)\,
g_j(t^\prime)
\label{eq:particleqnew}
\end{equation}
and
\begin{equation}
\frac{df_j(t)}{dt} = - \int_0^t dt^\prime\, R_j(t-t^\prime)\,
f_j(t^\prime),
\label{eq:holeqnew}
\end{equation}
respectively, for $t>0$. The computational advantage of the new
equations over the original ones is obvious. For a given $t$, the
integration runs over previous times only. Therefore, one can use the
equations to iterate the initial condition forward in time to a
certain given time $t>0$. The error made in taking the retarded
functions instead of the time-ordered one is measured by the parameter
$|V^{jm}_{kl}|/\varepsilon_j$. In the region of the expected
transition, $\varepsilon_j \sim \Delta \sqrt{g}$, this parameter is
small. We believe therefore that the retardation ansatz,
Eqs. (\ref{eq:particleqnew}) and (\ref{eq:holeqnew}), carries the same
physical information as the self-consistent approximation,
Eq. (\ref{eq:selfenergy}).

%%%%%%%%%%%%%%%%%%%%%%%%%%%%%%%%%%%%%%%%%%%%%%%%%%%%%%%%%%%%%%%%%%%%%%%%%%

\subsection{Inverse participation ratio and real-time quasiparticle
decay}

Let us discuss  characterization of the transition between the
extended and localized states in the Fock space by the real-time
single-particle Green's function method. Going back to
Eq. (\ref{eq:GF}), for $t>0$, one finds
\begin{eqnarray}
G_j(t) & = & -i \sum_\alpha \left\langle 0 \left| c_j(t) \right|
\alpha \right\rangle \left\langle \alpha \left| c^\dagger_j(0) \right|
0 \right\rangle \nonumber \\ & = & -i \sum_\alpha e^{it(E_0 -
E_\alpha)}\, 
\Big| 
\left\langle 0 \left| c_j \right| \alpha \right\rangle
\Big|^2 
%\left\langle \alpha \left| c^\dagger_j \right| 0 \right\rangle ,
\end{eqnarray}
where $|\alpha\rangle$ denotes an eigenstate of the full Hamiltonian,
$H$, with corresponding eigenenergy $E_\alpha$. As a result:
\begin{eqnarray}
\left| G_j(t) \right|^2 & = & \sum_\alpha \left| \left\langle 0 \left|
c_j \right| \alpha \right\rangle \right|^4 + 2 \sum_{\alpha>\beta}
\left| \left\langle 0 \left| c_j \right| \alpha \right\rangle
\right|^2 \left| \left\langle 0 \left| c_j \right| \beta \right\rangle
\right|^2 \nonumber \\ & & \times \cos \left[(E_\alpha - E_\beta)t
\right]\, .
\label{eq:absGF}
\end{eqnarray}
Upon ensemble averaging, the oscillating term disappears. Indeed,
since $E_\alpha - E_\beta$ behaves as a random variable, at long
enough times  the average of the oscillating term in
Eq. (\ref{eq:absGF}) rapidly goes to zero. One thus obtains
\begin{equation}
I_j \equiv \left. \overline{\left| G_j(t) \right|^2}
\right|_{t\rightarrow\infty} = \overline{\sum_\alpha \left|
\left\langle 0 \left| c_j \right| \alpha \right\rangle \right|^4}\, ,
\label{eq:IPR}
\end{equation}
where $I_j$ is the inverse participation ratio (IPR) for the Fock
state $c_j^\dagger|FS\rangle$. In other words, $P_j = 1/I_j$ is the
number of many-body eigenstates that overlap with the quasiparticle
state $c_j^\dagger|FS\rangle$. If a quasiparticle is localized in Fock
space, $P_j \approx 1$, while in the extended case $P_j \gg 1$. Thus,
for extended states $I_j \rightarrow 0$ in the thermodynamic limit. In
the later case one would expect the transition from localized and
extended states to be marked by an abrupt jump in $I_j$ as some
parameter (the excitation energy, $\varepsilon_j$, interaction
strength, $\lambda$, or the dimensionless conductance, $g$) is varied.

%%%%%%%%%%%%%%%%%%%%%%%%%%%%%%%%%%%%%%%%%%%%%%%%%%%%%%%%%%%%%%%%%%%%%%%%%%

\section{Numerical simulations}
\label{sec:III}

%%%%%%%%%%%%%%%%%%%%%%%%%%%%%%%%%%%%%%%%%%%%%%%%%%%%%%%%%%%%%%%%%%%%%%%%%%

Equations (\ref{eq:particleq}) and (\ref{eq:holeq}) have obvious
advantages over Eq. (\ref{eq:dyson}). Knowing $g_j(t)$ and $f_j(t)$
for all $j$ and at all times $t \le t_0$ allows one to determine
$g_j(t_0+\delta t)$ and $f_j(t_0+\delta t)$ by iteration, for
sufficiently small $\delta t$. Thus, one can calculate the Green's
function at any time starting from the initial conditions $g_j(0) =
f_j(0) = 1$ and $\dot{g}_j(0) = \dot{f}_j(0) = 0$ (the latter follow
from the integro-differential equations because the kernels are
non-singular). Since no inversion or diagonalization is involved, one
can treat rather large systems. In the present method the time cost
increases with the number $M$ of single-particle basis states. For the
unconstrained summations in Eqs. (\ref{eq:particlekernel}) and
(\ref{eq:holekernel}) the time it takes to go through $K$ iterations
goes as $K^2 M^4$ when $N \lesssim M$. Nevertheless, this time can be
substantially shorter than the diagonalization of a full many-body
Hamiltonian matrix defined over the same $M$ single-particle basis
states.

The weakest point of the iterative calculation is the appearance of
spurious numerical instabilities caused by the non-linearity of the
coupled equations. However, we have found a good balance between $K$,
$N$, and $g$ (for simplicity, we fixed $\lambda = 1$). When the
interactions are strong (small $g$) and instabilities are more
pronounced, the functions $g_j(t)$ and $f_j(t)$ decay faster and,
despite $\delta t$ being small (as compared to the Heisenberg time
$t_H = 1/\Delta$), $K$ is not exceedingly large. If the interactions
are weak (large $g$), the decay is slow; we thus need to keep
iterating up to $t\gg t_H$ to achieve the stationary state implied by
Eq. (\ref{eq:IPR}). However, in this case instabilities nearly do not
occur and we can use larger $\delta t$. The number $K$ was thus
approximately the same for all values of $g$.

As pointed out in Ref. \onlinecite{leyronas99}, one can simplify the
numerical calculations without sacrificing the physics by restricting
the number of connected states in Fock space. In the so-called layer
model,\cite{georgeot97} all non-diagonal matrix elements $V_{jk}^{lm}$
except those with $j+k=l+m$ are set to zero. This model provides a
good approximation when the typical non-zero matrix element is much
smaller than $\Delta$. For the layer model the number of total
operations after $K$ iterations decreases to $K^2 M^3$.

In the numerical simulations we also assume that the single-particle
energy levels are random variables uniformly distributed in a sequence
of intervals
\begin{equation}
\varepsilon_j \in [\Delta(j - 1/2); \Delta(j + 1/2)],
\end{equation}
$j = 1,2, \ldots M$.\cite{leyronas99} This has the clear advantage of
providing some degree of level repulsion, consistent with Gaussian
orthogonal (time-reversal symmetric) ensemble of random matrices, but
without the inconvenience of a non-uniform level density. The number
$N$ of states with $i\le 0$ defines the number of particles present in
the Fermi sea. The matrix elements of the interaction have to obey
certain symmetries, namely,
\begin{equation}
\begin{array}{ll}
V_{kl}^{jm} = V^{kl}_{jm} & \mbox{(hermiticity)} \\ 
%V_{kl}^{jj} = V^{kk}_{jm} = 0 & \mbox{(only off-diagonal terms)} \\ 
V_{kl}^{jm} = - V_{kl}^{mj} = - V_{lk}^{jm} & \mbox{(fermionic
character)}.
\end{array}
\end{equation}
Each independent matrix element is chosen from a Gaussian distribution
with zero average and $\delta V = \Delta/g$ standard deviation. Notice
that, after setting $\lambda=1$, the interaction strength is
controlled by the dimensionless conductance $g\gg 1$. The numerical
algorithm solves the integro-differential equation by iteration
(Cauchy problem), as in simulations of the time evolution of dynamical
systems.

%%%%%%%%%%%%%%%%%%%%%%%%%%%%%%%%%%%%%%%%%%%%%%%%%%%%%%%%%%%%%%%%%%%%%%%%%%

\subsection{Results}

Below we present the numerical results obtained for the layer model
with 50 particles and 150 basis state and various values of $g$.

\begin{figure}
\epsfxsize=7cm\epsfbox{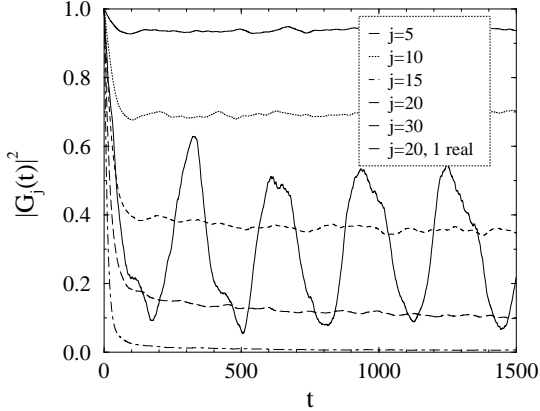}
\caption{Decay of the average return probability $\overline{\left|
G_j(t) \right|^2}$ for $g=100$ and various values of
$\varepsilon_j$. The solid line corresponds to a single realization
for $j=20$. Time is given in units of Heisenberg time, $t_H$.}
\label{fig1}
\end{figure}

Figure \ref{fig1} shows the averaged return probability
$\overline{\left| G_j(t) \right|^2}$ as a function of time $t$ for
various initial single-particle excited states $j$ when $g=100$. For
comparison, we have also plotted the same quantity taken for a single
realization (no averaging). Notice that in the latter case the return
probability decays rapidly and then oscillates strongly. The averaging
over different realizations suppresses such oscillations at large
times, yielding a saturation at the inverse participation ratio value
$I_j$. The higher the initial state energy $\varepsilon_j$, the lower
the inverse participation ratio, showing an increasing coupling to the
many-body eigenstates.

%%%%%%%%%%%%%%%%%%%%%%%%%%%%%%%%%%%%%%%%%%%%%%%%%%%%%%%%%%%%%%%%%%%%%%%%%%%%%%

%%%%%%%%%%%%%%%%%%%%%%%%%%%%%%%%%%%%%%%%%%%%%%%%%%%%%%%%%%%%%%%%%%%%%%%%%%%%%%

\begin{figure}
\epsfxsize=8cm\epsfbox{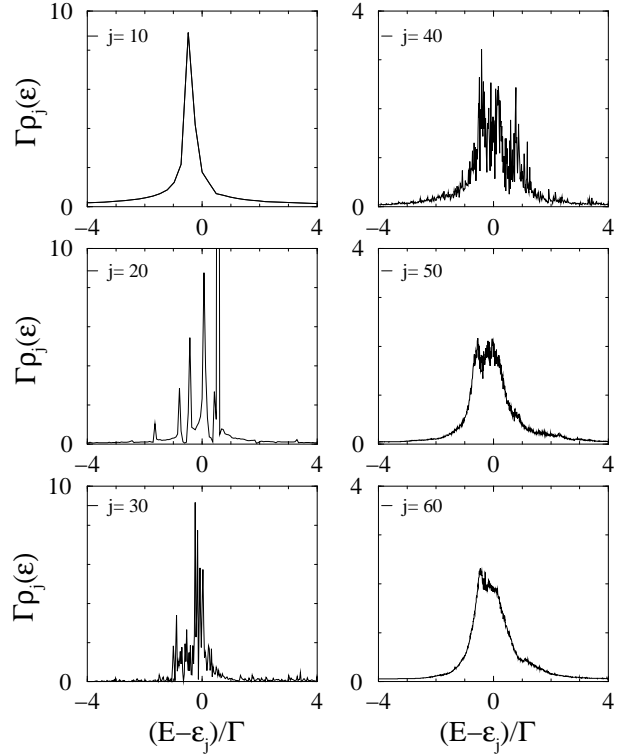} 
\caption{Local density of states $\rho_j(E) = - \frac{1}{\pi} {\rm Im}
\tilde{G}_j(E)$ for $g = 100$ but different quasiparticles energies
$\varepsilon_j$. $\Gamma$ is the Breit-Wigner width expected from the
golden rule prediction.}
\label{fig2}
\end{figure}

Figure \ref{fig2} shows the local density of states (LDOS)
$\rho_j(E)=- \frac{1}{\pi} {\rm Im} \tilde{G}_j(E)$ for a given
realization of the system, but different values of $j$. This quantity
characterizes the spreading of the single-particles states over the
many-body eigenstates. One may observe three distinct regimes as the
initial state energy is increased. At low $j$ the initial state
couples only to one or a few many-body eigenstates, thus corresponding
to a state localized in Fock space. As the energy $\varepsilon_j$
increases, the initial state begins to couple to a larger number of
many-body eigenstates; the resulting LDOS shows a ``fractal''
structure, with multiple peaks located near zero energy. As
$\varepsilon_j$ increases further, the initial single-particle state
overlaps to a much large number of many-body states, resulting in the
merging of peaks and a rather broad LDOS. It is important to remark
that upon ensemble averaging, $\rho_j(E)$ always develops into a
Breit-Wigner distribution (Lorentzian) regardless of the value of
$j$. Only the width of the distribution depends on the initial state
energy. Using the GR, one can estimate this width to be $\Gamma_j =
\frac{1}{3}\pi \Delta (j/g)^2$.\cite{sivan94b}

\begin{figure}
\epsfxsize=8cm\epsfbox{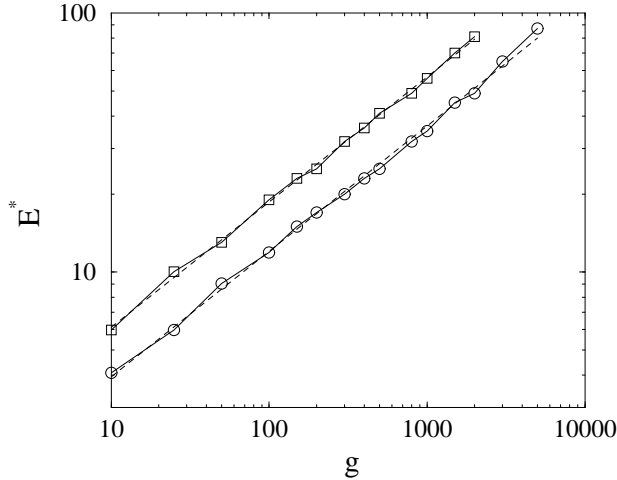} 
\caption{Delocalization energy $E^{\ast}$ for which $P_j= 2$
(circles) or $P_j = 10$ (squares) as a function of $g$. The dotted
lines show the fits for $E^\ast\sim \sqrt{g}$.}
\label{fig2'}
\end{figure}

According to the CT model predictions,\cite{altshuler97} the
transition between localized and extended states in Fock space occurs
at energy $E^\ast \approx \Delta \sqrt{g/\ln g}$. This implies a
delocalization threshold at $E^\ast/\Delta \approx 4.7 $ for $g =
100$. However, Fig. \ref{fig2} suggests instead that quasiparticles
survive for values of $\varepsilon_j$ up to $10$. In order to
determine the position of the delocalization threshold, we have
adopted the following criterion. A single-particle state is considered
well defined when its coupling to many-body states is very weak and
the corresponding participation number $P_j$ should not exceed some
small value (of the order of unity). Thus, we have estimated $E^\ast$
as the energy necessary for $P_j$ to reach a fixed value (for
instance, 2). The resulting scaling behavior with respect to $g$ is
shown in Fig. \ref{fig2'} and is accurately described by $E^\ast \sim
\Delta \sqrt{g}$. We also obtain the same scaling fixing $P_j
=10$. The difference between the square root scaling and the CT
prediction is likely to be linked to the assumption of constant
coordination number made in Ref. \onlinecite{altshuler97}. As in the
real situation, the branching number at each generation varies in the
numerical simulations; eventually, it decreases abruptly when the
number of particle-hole states is of the order of the number of
particles in the system. As it was suggested in
Ref. \onlinecite{mirlin97}, taking into account the decreasing of the
branching number with increasing the generation shifts the threshold
energy to higher levels. This agrees with our numerical data.

%%%%%%%%%%%%%%%%%%%%%%%%%%%%%%%%%%%%%%%%%%%%%%%%%%%%%%%%%%%%%%%%%%%%%%%%%%%%%%

%%%%%%%%%%%%%%%%%%%%%%%%%%%%%%%%%%%%%%%%%%%%%%%%%%%%%%%%%%%%%%%%%%%%%%%%%%%%%%

\begin{figure}
\epsfxsize=8cm\epsfbox{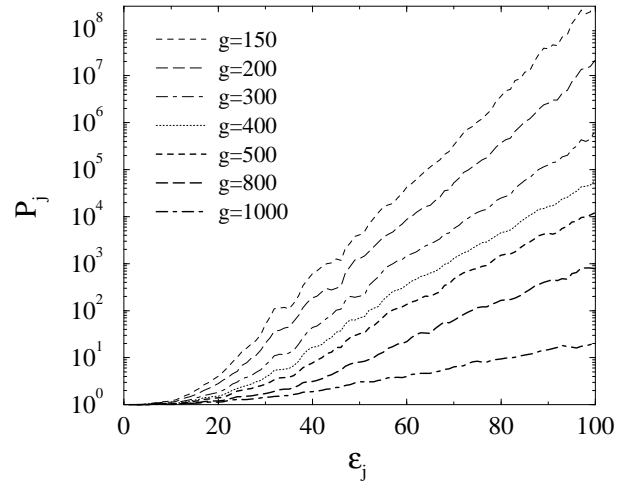} 
\caption{Participation ratio as function of energy $\varepsilon_j$ for
different values of $g$.}
\label{fig3}
\end{figure}

Following Ref. \onlinecite{mejia-monasterio98}, we attempt to
characterize the point of transition between the different regimes by
plotting the average participation number $P_j$ as a function of $j$
(energy) for different values of $g$. The result is shown in
Fig. \ref{fig3}. The absence of cusps or shoulders in the graphs is a
clear indication that no {\it sharp} transition between localized and
extended states in Fock space takes place. Instead, the data is
consistent with a continuous crossover, as previous exact
diagonalization of small systems indicated.\cite{mejia-monasterio98}
We have checked that the same behavior holds for systems twice as
large as well.

%%%%%%%%%%%%%%%%%%%%%%%%%%%%%%%%%%%%%%%%%%%%%%%%%%%%%%%%%%%%%%%%%%%%%%%%%%%%%%

%%%%%%%%%%%%%%%%%%%%%%%%%%%%%%%%%%%%%%%%%%%%%%%%%%%%%%%%%%%%%%%%%%%%%%%%%%%%%%

\begin{figure}
\epsfxsize=8cm\epsfbox{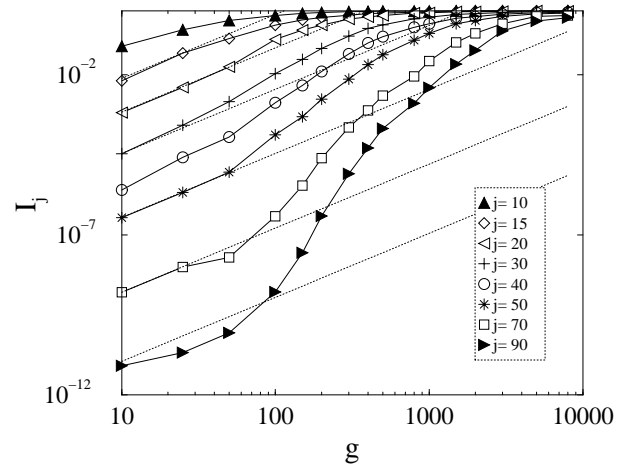}
\caption{Inverse participation ratio $I_j$ as function of the
dimensionless conductance $g$ for different values of the energy
$\varepsilon_j$. The dotted lines indicate the Fermi golden rule
prediction.}
\label{fig4}
\end{figure}

In the fully delocalized, ergodic regime, quasiparticles couple to all
energetically allowed many-body states. A way to characterize this
regime is to fit the IPR $I_j$ by the GR prediction, namely, $I_j
\approx {\rm min}(1,\delta /\Gamma)$, where $\delta$ is the mean
many-body level spacing at a given energy. Since $\delta /\Gamma
\propto g^2$,\cite{sivan94b} one expects a quadratic dependence of
$I_j$ with $g$. In Fig. \ref{fig4} we present the IPR $I_j$ as
function of $g$ for different values of $j$. It is clearly visible
that there is a faster than quadratic increase in the IPR for high
values of $j$. The quadratic behavior is restored as the interaction
strength grows ($g$ decreases). The failure of the GR prediction at
large $g$ occurs as one crosses a certain critical value, putting a
given energy $\varepsilon_j$ into the non--ergodic regime, namely,
$\varepsilon_j < E^{\ast\ast} (g)$. Although less pronounced, a
similar behavior was observed for smaller systems in
Ref. \onlinecite{leyronas99}.

While Fig. \ref{fig4} yields some information about the region where
the IPR deviates from the GR prediction, it does not make clear where
exactly ergodicity is broken. To quantify that Leyronas {\it et
al.}\cite{leyronas00} introduced a rescaled form of the IPR, $F = - y
\ln I_j$, where $x = (\varepsilon/g \Delta) \ln g$ is the rescaled
dimensionless conductance and $y = g (\Delta/\varepsilon)^{3/2}$ is
the rescaling factor.\cite{georgeot97} This quantity has the advantage
of illustrating both extreme situations of localized and ergodic
(delocalized) regimes in a much clearer way than $I_j$ itself. For
$x<1$ states are localized in Fock space and one can expand $F$ in
powers of $x$ for a fixed $\varepsilon_j$;
Ref. \onlinecite{silvestrov98} argues that the transition to the
ergodic regime occurs for values of the parameter $x \approx 1$. In
fact, at $x\gg 1$, the small-$x$ perturbation theory diverges and $F$
is expected to follow a behavior that can be deduced from the GR.

%%%%%%%%%%%%%%%%%%%%%%%%%%%%%%%%%%%%%%%%%%%%%%%%%%%%%%%%%%%%%%%%%%%%%%%%%%%%%%

%%%%%%%%%%%%%%%%%%%%%%%%%%%%%%%%%%%%%%%%%%%%%%%%%%%%%%%%%%%%%%%%%%%%%%%%%%%%%%

%%%%%%%%%%%%%%%%%%%%%%%%%%%%%%%%%%%%%%%%%%%%%%%%%%%%%%%%%%%%%%%%%%%%%%%%%%%%%%

%%%%%%%%%%%%%%%%%%%%%%%%%%%%%%%%%%%%%%%%%%%%%%%%%%%%%%%%%%%%%%%%%%%%%%%%%%%%%%

\begin{figure}
\epsfxsize=8cm\epsfbox{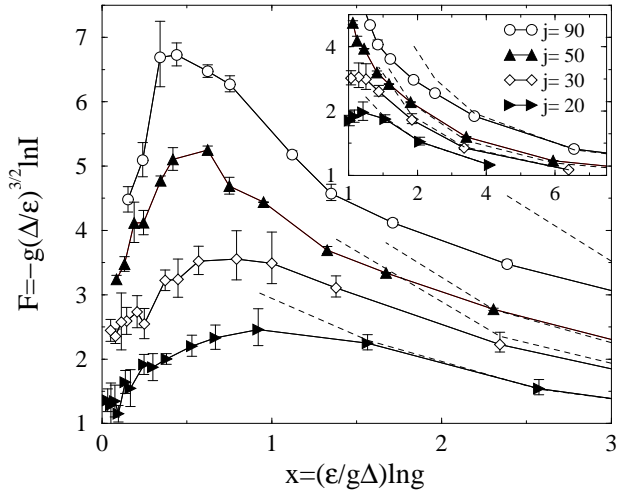}
\caption{Rescaled inverse participation ratio as function of the
effective interaction strength $x$ at different energies. The dashed
lines are the golder rule prediction. The inset shows in the detail
the large-$x$ behavior.}
\label{fig5}
\end{figure}

In Fig. \ref{fig5} we have plotted $F$ as function of the rescaled
dimensionless conductance $x$. Our results agree with those presented
in Ref. \onlinecite{leyronas00} for small energies. However, as the
energy is increased, three regimes (instead of two, as seen in
Ref. \onlinecite{leyronas00}) clearly appear. For large values of $g$
or small $x$ ($x<1$), where the perturbative calculation holds, the
maximum of the function $F$ shifts towards $x=0$, as $\varepsilon_j$
is increased. This may be interpreted as the appearance of higher than
linear contributions to the perturbative expansion. Another notable
behavior of the graphs at high $\varepsilon_j$ is the absence of
matching to the GR prediction for an extended range of $x$ (when
$x>1$). For instance, when $j = 90$, the GR prediction agrees with the
numerical data only at $x>4$. Thus, our results indicate existence of
a range of values of $g$ where there is neither localization nor
ergodicity (full delocalization). This range grows in size at higher
energies.

%%%%%%%%%%%%%%%%%%%%%%%%%%%%%%%%%%%%%%%%%%%%%%%%%%%%%%%%%%%%%%%%%%%%%%%%%%%%%%

%%%%%%%%%%%%%%%%%%%%%%%%%%%%%%%%%%%%%%%%%%%%%%%%%%%%%%%%%%%%%%%%%%%%%%%%%%%%%%

%%%%%%%%%%%%%%%%%%%%%%%%%%%%%%%%%%%%%%%%%%%%%%%%%%%%%%%%%%%%%%%%%%%%%%%%%%

\section{Discussion and conclusions}
\label{sec:IV}
 
%%%%%%%%%%%%%%%%%%%%%%%%%%%%%%%%%%%%%%%%%%%%%%%%%%%%%%%%%%%%%%%%%%%%%%%%%%

Using an approximate method in the time domain we have calculated the
single-particle Green's function and extracted information about the
quasiparticle decay rate. The method allows us to study larger systems
than previously treated in the literature. At low energies our results
are comparable to those obtained from direct diagonalization in small
systems. At higher energies, however, we are able to identify three
different regimes as function of the effective interaction
strength. Furthermore, we have found that the delocalization threshold
occurs at energies higher than the prediction of the CT mapping,
$\Delta \sqrt{g/\ln g}$; from our data, we estimate $E^\ast \sim
\Delta \sqrt{g}$.

It is interesting to notice that a sequence of three different regimes
in the absence of sharp transitions also occurs for quantum systems
whose classical dynamics corresponds to the Komolgorov-Arnold-Moser
(KAM) regime.\cite{georgeot97} Indeed, drawing an analogy between the
present problem and the KAM theorem of classical mechanics is very
tempting. In that sense, the problem of a quasiparticle decay in a
finite interacting system would be interpreted as a type of {\it
soft}, interaction-induced crossover to chaos\cite{leyronas00} rather
than an abrupt transition for a particular value of $\varepsilon_j$.

Within this interpretation, the scaling behavior seen in the
perturbative expansion around the localized state\cite{silvestrov97}
corresponds to the theory of a regular classical system subjected to
small symmetry breaking (potentially chaotic) perturbation. The radius
of convergence of such theory breaks down at finite perturbation
strength, indicating the onset of mixed dynamics (destruction of the
first phase space tori). Beyond the convergence radius, soft chaos is
present, but there is no complete ergodicity yet. Indeed, our data for
the IPR follow a similar behavior for intermediate values of the
interaction strength $g$. The KAM analogy would be completed by
connecting the onset of ergodicity in classical phase space (at
sufficiently strong perturbations) to onset of the regime where the GR
prediction for the IPR apply.

However, it is important to remark that we have not attempted to
construct in any form a classical counterpart of the quantum many-body
system.\cite{saraceno} Also, our approach does not give any
information about the nature or the statistical fluctuations of the
many-body excited states. Thus, we cannot confirm whether the
quasiparticle decay problem and the KAM theorem are deeply related, or
they just provide coincident regimes.

Recently, a few theoretical predictions have been made about the
dynamics that governs the decaying
process.\cite{izrael01,silvestrov01} So far, no experimental or
numerical simulations have corroborate these predictions. We plan to
investigate this problem using the method here introduced.

%%%%%%%%%%%%%%%%%%%%%%%%%%%%%%%%%%%%%%%%%%%%%%%%%%%%%%%%%%%%%%%%%%%%%%%%%%

\section{Acknowledgments}

We are indebted to L. S. Levitov for suggesting the real-time approach
used in this work. Stimulating discussions with B. L. Altshuler and
C. H. Lewenkopf are acknowledged. A.M.F.R. and E.R.M. thank the
Brazilian agencies CNPq, FAPERJ, and PRONEX for financial support.
A.K. was partially supported by the BSF grant No. 9800338.

%%%%%%%%%%%%%%%%%%%%%%%%%%%%%%%%%%%%%%%%%%%%%%%%%%%%%%%%%%%%%%%%%%%%%%%%%%

%%%%%%%%%%%%%%%%%%%%%%%%%%%%%%%%%%%%%%%%%%%%%%%%%%%%%%%%%%%%%%%%%%%%%%%

\end{multicols}

\end{document}